\documentclass[aps, twocolumn, superscriptaddress, nobalancelastpage]{revtex4-1}

\usepackage{amsfonts}
\usepackage{graphicx}
\usepackage{tabularx}
\usepackage{times}
\usepackage{hyperref}
\usepackage{enumerate}

\usepackage[usenames]{xcolor}
\hypersetup{
	colorlinks,
	linkcolor={blue!50!black},
	citecolor={blue!50!black},
	urlcolor={blue!80!black}
}
\begin{document}
\title{Machine Learning in a data-limited regime:\\Augmenting experiments with synthetic data uncovers order in crumpled sheets}

\date{\today}

\author{Jordan Hoffmann}
\affiliation{John A.~Paulson School of Engineering and Applied Sciences, Harvard University, Cambridge, MA 02138, USA}
\author{Yohai Bar-Sinai}
\email[Corresponding author: ]{ybarsinai@gmail.com}
\affiliation{John A.~Paulson School of Engineering and Applied Sciences, Harvard University, Cambridge, MA 02138, USA}
\author{Lisa Lee}
\affiliation{John A.~Paulson School of Engineering and Applied Sciences, Harvard University, Cambridge, MA 02138, USA}
\author{Jovana Andrejevic}
\affiliation{John A.~Paulson School of Engineering and Applied Sciences, Harvard University, Cambridge, MA 02138, USA}
\author{Shruti Mishra}
\affiliation{John A.~Paulson School of Engineering and Applied Sciences, Harvard University, Cambridge, MA 02138, USA}
\author{Shmuel M. Rubinstein}
\email[Corresponding author: ]{shmuel@seas.harvard.edu}
\affiliation{John A.~Paulson School of Engineering and Applied Sciences, Harvard University, Cambridge, MA 02138, USA}
\author{Chris H. Rycroft}
\affiliation{John A.~Paulson School of Engineering and Applied Sciences, Harvard University, Cambridge, MA 02138, USA}
\affiliation{Computational Research Division, Lawrence Berkeley Laboratory, Berkeley, CA 94720, USA}

\begin{abstract}
Machine learning has gained widespread attention as a powerful tool to identify structure in complex, high-dimensional data. However, these techniques are ostensibly inapplicable for experimental systems where data is scarce or expensive to obtain. Here we introduce a strategy to resolve this impasse by augmenting the experimental dataset with synthetically generated data of a much simpler sister system. Specifically, we study spontaneously emerging local order in crease networks of crumpled thin sheets, a paradigmatic example of spatial complexity, and show that machine learning techniques can be effective even in a data-limited regime. This is achieved by augmenting the scarce experimental dataset with inexhaustible amounts of simulated data of rigid flat-folded sheets, which are simple to simulate and share common statistical properties. This significantly improves the predictive power in a test problem of pattern completion and demonstrates the usefulness of machine learning in bench-top experiments where data is good but scarce.
\end{abstract}

\maketitle

\section*{Introduction}

Machine learning is a versatile tool for data analysis that has permeated applications in a wide range of domains~\cite{Yann2015}. It has been particularly well suited to the task of mining large datasets to uncover underlying trends and structure, enabling breakthroughs in areas as diverse as speech and character recognition~\cite{Mohamed2009,Dahl2011,Zweig2013,MNIST}, medicine~\cite{Shameer2016}, games~\cite{Mnih2013,Silver2017}, finance~\cite{Heaton2016}, and even romantic attraction~\cite{Joel2017}. The prospect of applying machine learning to research in the physical sciences has likewise gained attention and excitement. Data-driven approaches have been successfully applied to data-rich systems such as classifying particle collisions in the LHC~\cite{Bhimji2017,Baldi2014}, classifying galaxies~\cite{Banerji2009}, segmenting large microscopy datasets~\cite{Sommer2011,GoogleFundus} or identifying states of matter~\cite{Carrasquilla2017,Spellings2018}. Machine learning has also enhanced our understanding of soft-matter systems: In a recent series of works, Cubuk, Liu, and collaborators have used data-driven techniques to define and analyze a novel ``softness'' parameter governing the mechanical response of disordered, jammed systems~\cite{cubuk,Sussman2016,Schoenholz2016}.

All examples cited above address experimentally, computationally, or analytically well-developed scientific fields supplied by effectively unlimited data. By contrast, many systems of interest are characterized by scarce or poor-quality data, a lack of established tools, and a limited data acquisition rate that falls short of the demands of effective machine learning. As a result, the applicability of machine learning to such systems is problematic and would require additional tools. This would potentially be of high value to the experimental physics community and would require novel ways of circumventing the data limitations, either experimentally or computationally. 

In this manuscript, we study crumpling and the evolution of damage networks in thin sheets as a test case for machine-learning-aided science in complex, data-limited systems that lack a well established theoretical, or even a phenomenological, model.

Crumpling is a complicated and poorly understood process: As a thin sheet is confined to a small region of space, stresses spontaneously localize into one-dimensional regions of high curvature~\cite{BenAmar1997,Witten2007,Aharoni2010}, forming a damage network of sharp creases (Fig.~1B) that can be classified according to the sign of the mean curvature: creases with positive and negative curvature are commonly referred to as valleys and ridges, respectively. Previous works on crumpled sheets have established clear and robust statistical properties of these damage networks. For example, it has been shown that the number of creases at a given length follows a predictable distribution~\cite{Andresen2007}, and the cumulative amount of damage over repeated crumpling is described by an equation of state~\cite{Omer2018}. However, these works do not account for spatial correlations, which is the structure we are trying to unravel.  The goal of this work is to learn the statistical properties of such networks by solving a problem of network completion: Separating the ridges from valleys, can a neural net be trained to accurately recover the location of the ridges, presented only with the valleys? For later use, we call this problem \emph{partial network reconstruction}.
The predominant challenge we are addressing here is a severe data limitation. As detailed below, we were unable to perform this task using experimental data alone. However, by augmenting experimental data with computer-generated examples of a simple sister system which is well understood, namely rigid flat-folding, we trained an appropriate neural network with significant predictive power.

\begin{figure}
	\centering
	\includegraphics[width=\columnwidth]{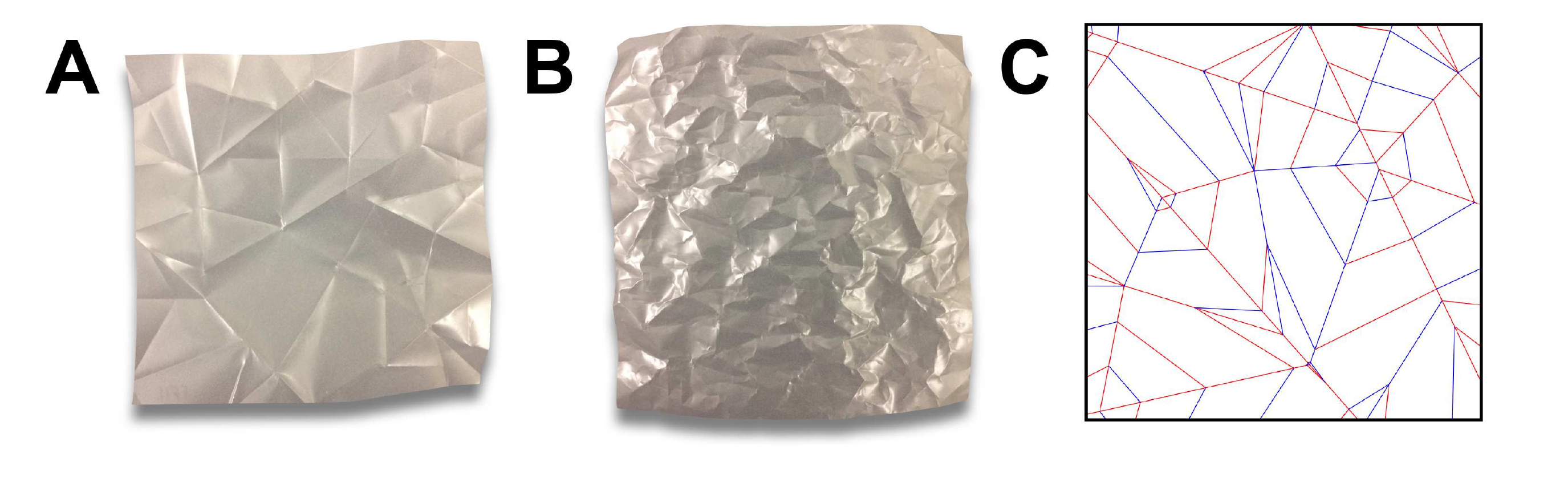}\label{fig:DSLR}
	\caption{
	\textbf{Examples of crease networks:} (A) A $10\textrm{~cm}\times 10\textrm{~cm}$ sheet of Mylar that has undergone a succession of rigid flat-folds. (B) A sheet of Mylar that has been crumpled. (C) A simulated rigid flat-folded sheet. The sheet has been folded 13 times. Ridges are colored red, and valleys are blue.}
\end{figure}

The primary dataset used in this work was collected for a previous crumpling study~\cite{Omer2018}, where the experimental procedures are detailed and are only reviewed here for completeness. $10\textrm{~cm}\times10\textrm{~cm}$ Mylar sheets are crumpled by rolling them into a 3~cm diameter cylinder and compressing them uni-axially to a specified depth within the cylindrical container, creating a permanent damage network of creasing scars embedded into the sheet. To extract the crease network, the sheet is carefully opened up and scanned using a custom-made laser profilometer, resulting in a topographic height map from which the mean curvature is calculated. The sheet is then successively re-crumpled and scanned between 4 and 24 times, following the same procedure. The curvature map is preprocessed with a custom algorithm based on the Radon transform (for details see Sec.~I in the Supplementary Information (SI)) to separate creases from the flat facets and fine texture in the data (Fig.~2A). The complete dataset consists of a total of 506 scans corresponding to 31 different sheets.

\section*{Results}
\paragraph*{Failures with only experimental data}
As stated above, the task we tried to achieve is partial network reconstruction: inferring the location of the ridges given only the valleys (Fig.~2A). Our first attempts were largely unsatisfactory and demonstrated little to no predictive power. Strategies for improving our results included subdividing the input data into small patches of different length scales, varying the network architecture, data representation, and loss function, and denoising the data in different ways. We approached variants of the original problem statement, trying to predict specific crease locations, distance from a crease, and changes in the crease network between successive frames. In all these cases our network invariably learned specific features of the training set rather than general principles that hold for unseen test data, a common problem known as over-fitting. The main culprit for this failure is insufficient data: the dataset of a few hundred scans available for this study is small compared to standard practices in machine learning tasks (for example, the problem of hand-written digit classification, \texttt{MNIST}, which is commonly given as an introductory exercise in machine learning, consists of 70,000 images~\cite{MNIST}). Moreover, as creases produce irreversible scars, images of successive crumples of the same sheet are highly correlated, rendering the effective size of our dataset considerably smaller. 

Over-fitting can be addressed by constraining the model complexity through insights from physical laws, geometric rules, symmetries, or other relevant constraints. Alternatively, it can be mediated by acquiring more data. Sadly, neither of these avenues is viable: current theory of crumpling cannot offer significant constraints about the structure or evolution of crease networks. Furthermore, adding a significant amount of experimental data is prohibitively costly: achieving a dataset of the size typically used in deep learning problems, say $10^4$ scans, would require thousands of lab hours, given that a single scan takes about ten minutes. Lastly, data cannot be efficiently simulated since, while preliminary work on simulating crumpling is promising~\cite{Rahul2013,Guo2018}, generating a simulated crumpled sheet still takes longer than an actual experiment. A different approach is needed.

\begin{figure}
	\centering
	\includegraphics[width=\columnwidth]{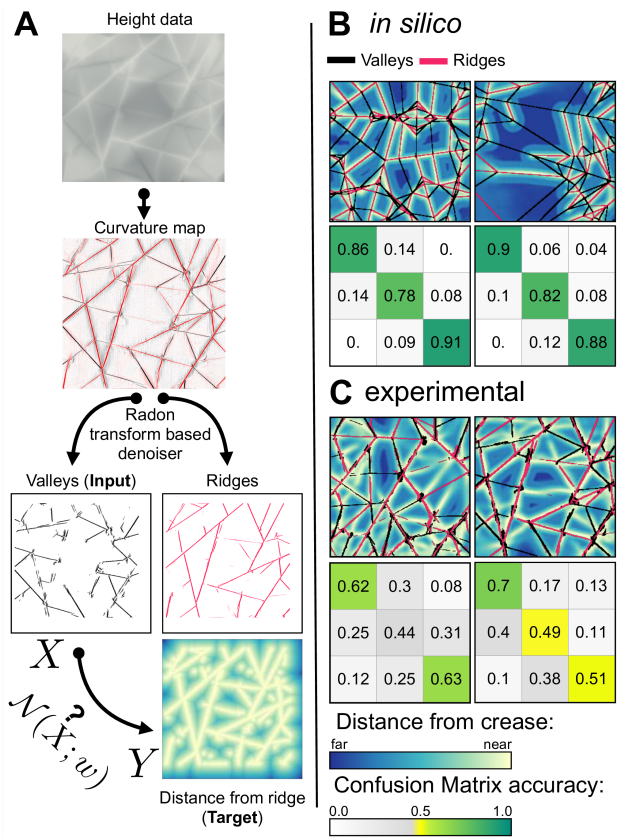}\label{fig:schematic}
	\caption{(A) \textbf{A schematic of the processing pipeline}. From the height map, a mean curvature map is calculated and denoised with a Radon-transform based method. Valleys (black) and ridges (red) are separated. The binary image of the valleys ($\mathbf{X}$) is the input to the neural network ($\mathcal{N}$). The distance transform of the binary image of the ridges is the target ($\mathbf{Y}$). Brighter colors represent regions closer to ridges. These color conventions are consistent through all figures in this paper. (B) Two samples of predictions on generated data. The true fold network is superimposed on the predicted distance map. It is seen that the true ridges (red) coincide perfectly with the bright colors, demonstrating strong predictive power. Below the predictions we show confusion matrices, with the nearest third of pixels, the middle third, and the furthest third. (C) Two predictions, and their corresponding confusion matrices, using the network trained on generated data (without noise) and applied to experimental scans.}
\end{figure}

\paragraph*{Turning to a sister system: rigid flat-folding}
An alternative strategy is to consider a reference system free from data limitations alongside the target system, with the idea that similarities between the target and reference systems allow a machine learning model of one to inform that of the other. This is similar to \emph{transfer learning}~\cite{pan2010survey}, but in this case rather than re-purpose a network, we supplement the training data with that of a reference system. In our case, a natural choice of such a system is a rigid flat-folded thin sheet, effectively a more constrained version of crumpling which is well understood. Rigid flat-folding is the process of repeatedly folding a thin sheet along straight lines to create permanent creases, keeping all polygonal faces of the sheet flat during folding. For brevity, we will henceforth omit the word ``rigid'' and refer simply to flat-folding.

Known rules constrain the structure of the flat-folded crease network: Creases cannot begin or terminate in the interior of a sheet---they must either reach the boundary or create closed loops; the number of ridge and valley creases that meet at each vertex differs by two (Maekawa's theorem); finally, alternating sector angles must sum to $\pi$ (Kawasaki's theorem)~\cite{Turner2015}. Given these rigid geometric rules, we expect partial network reconstruction of  rigid flat-folded sheets to be a much more constrained problem than that of crumpled ones.

However, while experimentally collecting flat-folding data is only marginally less costly than collecting crumpling data, simulating it on a computer is a straightforward task, which provides a dataset of a practically unlimited size. We wrote a custom code to do this using the \texttt{Voro++} library~\cite{rycroft09c} for rapid manipulation of the polygonal folded facets, as described in Sec.~II of the Supplemental Information. Typical examples are shown in Fig.~1C, Fig.~2B and Fig.~S1.

Having flat-folding as a reference system provides foremost a convenient setting for comparing the performance of different network architectures. The vast parameter space of neural networks requires testing different hyperparameters, loss functions, optimizers, and data representations with no standard method for finding the optimal combination. This problem is exacerbated when it is not at all clear where the failure lies: Is the task at all feasible? If so, is the network architecture appropriate? If so, is the dataset sufficiently large? Answering these questions with our limited amount of experimental data is very difficult. In contrast, for flat-folded sheets we are certain the task is feasible and our data is comprehensive, so experimentation with different networks is easier. Indeed, after testing many architectures, we identified a network capable of learning the desired properties of our data, reproducing linear features and maintaining even non-local angle relationships between features.

\paragraph*{Network structure} The chosen network is a modified version of the fully-connected \texttt{SegNet}~\cite{segnet2015} deep convolutional neural net.  As outlined in Fig.~2A, each crease network is separated into its valleys and ridges. The neural net, $\mathcal{N}$, is given as an input a binary image of the valleys, denoted $\bf X$ (``input'' in Fig.~2). 
The output of the network, $\mathcal{N}(\bf X)$, is the predicted distance transform of the ridges, $\bf Y$.
That is, for each pixel, $\bf Y$ is the distance to the nearest ridge pixel (``target'' in Fig.~2). Training is performed by minimizing the $L_2$ distance (the ``loss'') between the predicted distance transform, $\hat{\bf Y}=\mathcal{N}(\bf X)$ and the real one, 
\begin{equation}
L=\sum_{i} (\hat{Y}_{i} - Y_{i})^2\ ,
\label{eq:loss}
\end{equation}
where the summation index $i$ represents image pixels. The motivation for this choice of representation is that creases are sharp and narrow features, and therefore if we require $\mathcal N$ to predict the precise location of a crease, even slight inaccuracies would lead to vanishing gradients of $L$, making training harder. See \textit{Materials and Methods} below for full details of the implementation.

\paragraph*{In silico flat-folding}
For exclusively \textit{in silico} generated flat-folding data, the trained network performs partial network reconstruction with nearly perfect accuracy, as demonstrated in Fig.~2B: The agreement between the true location of the valleys (red lines) and their predicted location (bright colors) is visibly flawless. As a means of quantifying accuracy, we present the confusion matrices of the predicted and true output (Fig.~2B).

Confusion matrices are a common way to quantify classification errors, and since we are predicting the distance from a crease, the problem can be thought of as a classification problem: choosing some thresholds according to typical values of the distances, we can ask for each point in space whether it close to a crease, far from it, or at an intermediate distance. The confusion matrix measures what percentage of each class is correctly classified, and if not, what class it is wrongly classified as. We define three equal bins, based on the relative distance from the predicted ridges. The upper row in the matrix corresponds to pixels which are closest to ridges, and the lower row to farthest pixels. Similarly, the first and last column correspond to the closest and farthest predicted distances. Thus, the top left entry in the matrix  contains the probability of correctly predicting regions closest to a ridge, which is approximately 90\%.

Partial network reconstruction of \textit{in silico} flat-folded sheets is itself a non-trivial task requiring the knowledge of a complicated set of geometrical rules. Tasked to a human, inferring these rules from the data would require non-negligible effort in writing an explicit algorithm. The neural network, however, solves this problem with relative ease. 

\paragraph*{Experimental flat-folding}
As an intermediate step between \textit{in silico} flat-folding and experimental crumpling data, we next examine the performance of the neural network on experimental flat-folding scans. Fig.~2C reveals that the resulting prediction weakens by comparison, a consequence of noise present in experimental data that is absent from the \textit{in silico} samples. Noise occurs in the form of varying crease widths, fine texture, and missing creases that are undetected in image processing. In some cases, even the true creases that are missed during processing are correctly predicted, which also introduces error to our accuracy metric (see for example, the center of the second panel of Fig.~2C). While sufficient data of experimental flat-folding would likely allow the network to distinguish signal from noise, in our data-limited regime noise must be added to the generated \textit{in silico} data in order to help the network learn to accurately predict experimental scans and avoid over-fitting.

\begin{figure}
	\centering
	\includegraphics[width=\columnwidth]{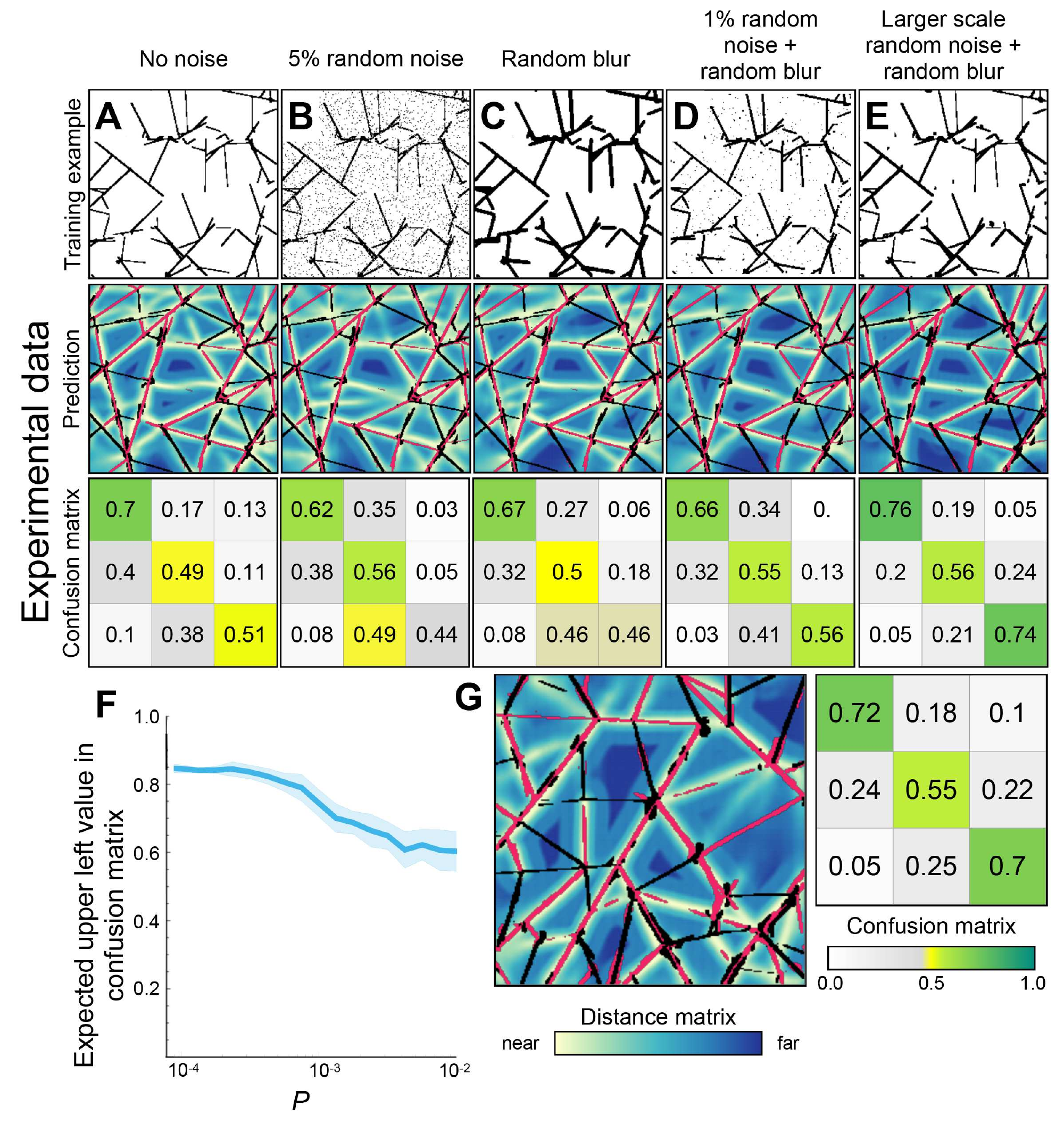}\label{fig:noise} 
	\caption{\textbf{Effect of noise type on prediction} (A)--(E) An example noised image (top), an example prediction (middle) and the corresponding confusion matrix (bottom) for different types of artificial noise. Noise types are described concisely in the title of each panel and complete specifications are given in \emph{Materials and Methods}. (F) The upper left value of the confusion matrix when each pixel of the near perfect prediction from Fig.~2B was randomly toggled with probability $P$ (G) The network from (E) applied on an additional experimental scan (from left panel of Fig.~2C). The average confusion matrix on all experimental scans is shown.}
\end{figure}

We examine the effect of adding several types of noise on the prediction accuracy on experimental input (Figs.~3A--E). We observe significant improvement and find that adding experimentally realistic noise (Fig.~3E) is more effective than toggling individual pixels randomly (Figs.~3B,D). We found that the noise type that leads to optimal training is to randomly add and remove patches of input that are approximately the same length scale as the noise in the experimental scans. We also find that it is important to provide input data with lines of variable width, to prevent the network from expecting only creases of a particular width. For complete details of the different noise properties, see \emph{Materials and Methods}. 

While the values in the confusion matrices in Fig.~3E might seem low, it is important to note that the metric used here is not trivial to interpret: it compares the $L_2$ distance from a distance map, which is particularly sensitive to noise since a localized noise speckle in a region remote from valleys perturbs a large region of space (essentially, of the size of its Voronoi cell). To gauge the effect of noise on the accuracy metric, we randomly toggle a fraction $P$ of pixels in an otherwise perfect flat-folding example and recompute the entries of its confusion matrix, as presented in Fig.~3F. With realistic noise levels, i.e.~$P\sim 10^{-3}$, we can expect accuracy values between 0.75 and 0.80 in the upper left and lower right entries of the confusion matrix, comparable to the values reported in Fig.~3E. That is, for experimental flat-folding we achieve accuracy levels that are comparable to what is expected for a perfect prediction with noisy preprocessing.

\paragraph*{Experimental crumpling}
For crumpling, we train the neural network using a combination of 30\% experimental crumpling and 70\% \emph{in silico} flat-folding data, that was noised as described above.  We also tried pre-training on in-silico data prior to training on crumpling data, but observed no improvement. Training on this combined dataset, the resulting predictions accurately reconstruct key features of the crease networks in crumpled sheets, that were not achieved in prior attempts. In Figure~4, we present predictions on entire sheets (Fig.~4A) as well as a few close ups on selected regions (Fig.~4B). The confusion matrices suggest that the network is often relatively accurate in predicting regions that are directly near a crease (upper left entry) as well as large open spaces (lower right entry), classifying such regions with 50\%--60\% accuracy. In addition, Fig.~S3 shows the prediction on each of 16 successive crumples of the same sheet held out from training. 

\begin{figure*}
	\centering
	\includegraphics[width=\textwidth]{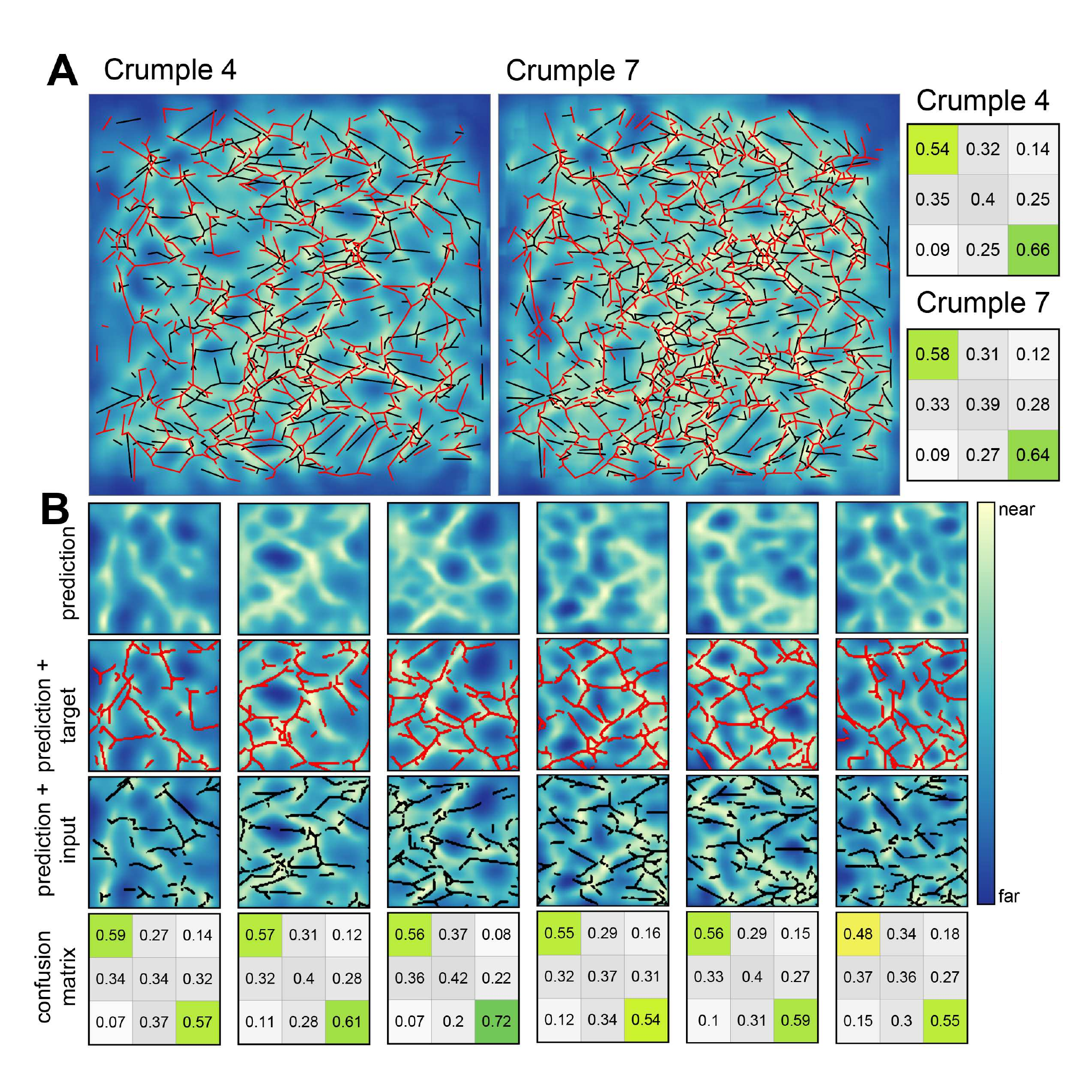}
	\label{fig:crump}
	\caption{\textbf{Predictions on crumpling}  (A) One sheet that was successively crumpled, shown after 4 and 7 crumpling iterations. Color code follows Fig.~2. (B) Close-ups on selected smaller patches from the same image, broken down to prediction, prediction and target, and prediction and input. }
\end{figure*}

\begin{figure*}
	\centering
	\includegraphics[width=16 cm]{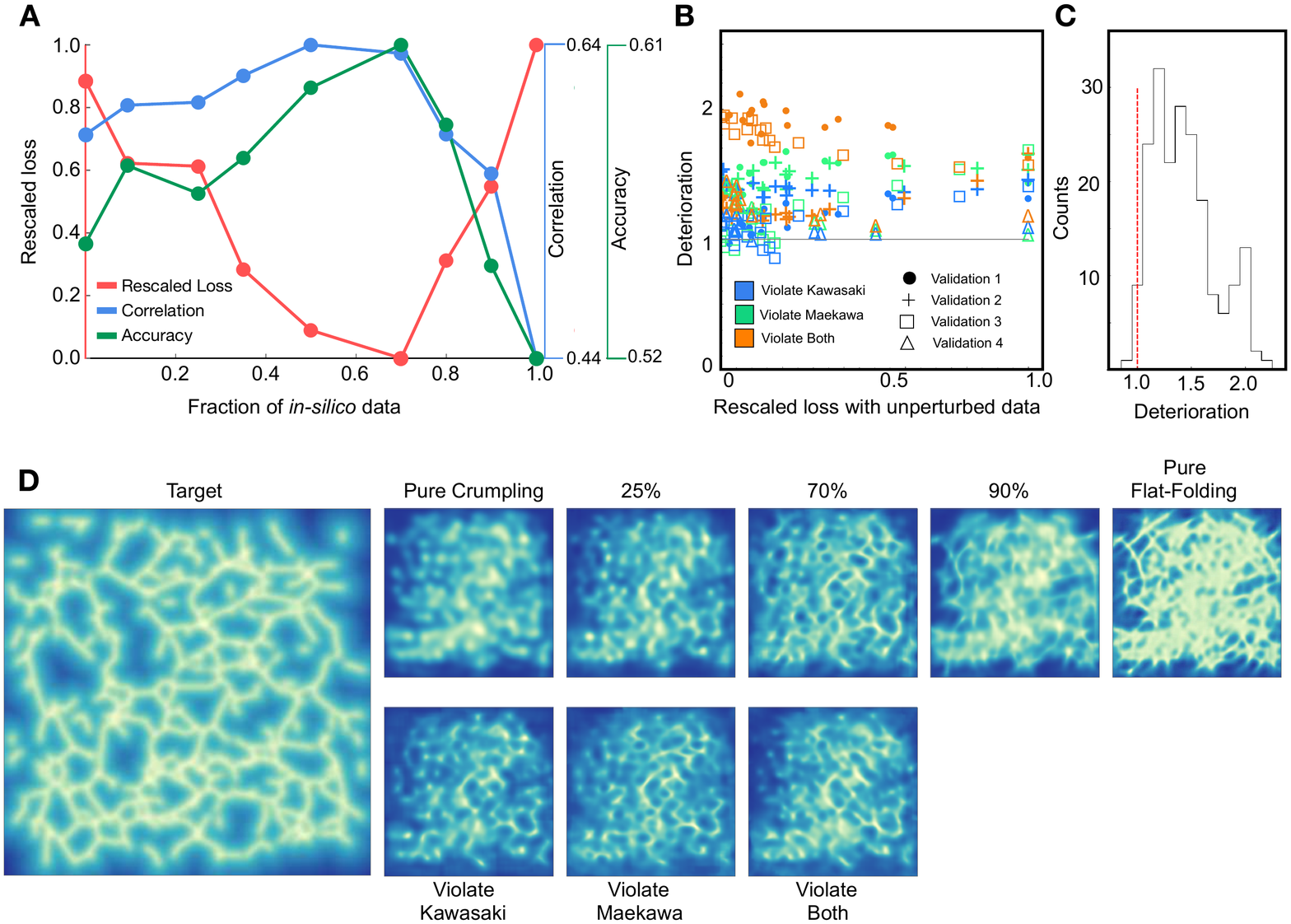}\label{fig:lambda}
	\caption{\textbf{Effect of fraction generated data} (A) Three quantifications of the predictive power of the model when trained on a varying amount of generated data and a constant amount of crumpling data. Strong predictive power corresponds to low loss (red) and large Pearson correlation and classification accuracy (blue and green respectively). 
		(B) The Deterioration (see text) for each sheet in the validation set, as a function of the rescaled loss.
		Colors correspond to different perturbations and marker styles to cross-validation sample. 
		It is seen that all tested perturbations lead to worse predictive power (above the gray reference line). The few points below the reference line occur at high crumple number, and low absolute loss.
		(C) Histogram of all points in panel~B. Values to the right of the red line corresponds to deterioration when using unphysical data. 
		(D) Example target and predictions for the various models considered in previous panels.
	}
\end{figure*}

The ratio of 70\% \emph{in silico} data was chosen since it provides optimal predictions, as shown in Fig.~5A. We  present three different metrics to quantify the predictive power: the $L_2$ loss of Eq.~\ref{eq:loss}, the Pearson correlation between the prediction and the target, and the average of the upper left and lower right of the confusion matrix (classification accuracy). We find that all accuracy metrics are optimized for training on 50\%-70\% \textit{in silico} data. It is also interesting to see in what way this affects the  prediction: In Fig.~5D we show that when trained solely on experimental data the neural network produces a blurred and indecisive prediction, while for 100\% flat-folding data the network predicts only unrealistic straight and long creases.

In addition to these metrics, one can compare the network's output to ``random'' network completion, i.e.~to a network that construes a pattern having the statistical properties of a crease network, but is only weakly correlated with the input image. Though a generative model for crease networks is not available, we can sample crease patterns from the experimental data and compare the predicted distance maps to those measured from such randomly selected samples. This is discussed in Sec.~V of the SI, where it is seen that our prediction for a given crease pattern is overwhelmingly closer to the truth than any sampled patch from other experiments (Fig.~S5).

\paragraph*{The similarity of flat-folding and crumpling}
These results demonstrate that augmenting the dataset with \emph{in silico} generated flat-folding data allows the network to discern some underlying geometric order in crease networks of experimental crumpling data. This suggests that the two systems share some common statistical properties, and it is interesting to ask how robust this similarity is. One may suspect that the main contribution of the \textit{in silico} data is merely having a multitude of intersecting straight lines, which are the main geometric feature that is analyzed, but that the specific statistics of these lines is not crucial.

As explained above, flat-folding networks are characterized by two theorems: Maekawa's theorem that constrains the curvatures (ridge/valley) of creases joined in each vertex, and Kawasaki's theorem which constrains the relative angles at vertices. We tested the sensitivity of our prediction to replacing the \emph{in silico} data used in training with crease networks that violate these rules: We obtained crease networks that violate Maekawa's theorem by taking flat-folding networks and randomly reassigning curvatures to each crease, and crease networks that violate Kawasaki's theorem by perturbing all vertex positions. Finally, we obtained crease networks that violate both rules by performing both perturbations simultaneously. Examples of perturbed networks are shown in Fig.~S6 of the Supplementary Information, with additional details about the perturbation process.

The effect is quantified in Fig.~5B and Fig.~5C. We define, for a given sheet, the ``Deterioration'' as the ratio between the loss of a network trained on 70\% experimental data and 30\% perturbed flat-folding data, to that of a network trained on the same ratio of experimental and unperturbed data.
It is seen that breaking the flat-folding rules leads to consistently worse performance, for all types of perturbations.

We cross-validated with 4 different experiments covering a total of 198 sheets. While for some small fraction ($<5\%$) of the sheets perturbed data has led to marginally better performance, this happened mostly in sheets with low loss and the improvement is negligible. On average, the network trained on perturbed data has a loss approximately 35\% higher than that of the network trained on unperturbed data. 

These results, namely that training on perturbed flat-folding networks led to inferior performance, again suggest a similarity between crumpled crease networks and flat-folded networks. We did not quantitatively study the detailed effect of the different kinds of perturbations -- i.e.~whether violating Kawasaki's rule, Maekawa's rule, or both, results in more or less accurate predictions. Instead, equipped with this physical insight, we propose to directly probe the statistical similarity with traditional methods by measuring vertex properties in crease networks, a study which will be reported elsewhere.

\section*{Discussion}
Experimental data is paramount to our understanding of the physical world. However, prohibitive data acquisition rates in many experimental settings require augmenting experimental data in order to draw meaningful conclusions. In particular, computer simulations now play a significant role in exploratory science; many experimental conditions can be accurately simulated to corroborate our understanding of empirical results.

Despite these advances, the simulation of certain phenomena is inhibited by insufficient theoretical knowledge of the system, or by demanding computational resources and development time. For crumpling, without a deeper understanding which would allow the use of simplified/reduced models, simulations require prohibitively small time steps, small domain discretization, or both~\cite{Rahul2013}. In this manuscript, we show that even with a small experimental training set, augmenting the dataset by computer generated, artificially noised data of flat-folding, salient features of the ridge network can be predicted from the surrounding valleys: the network successfully predicts the presence of certain creases, as well as their pronounced absence in certain locations (see Fig.~4B). Moreover, our results demonstrate a statistical similarity between flat-folding and crumpling, evidenced by the fact that when flat-folding data is replaced with data of similar geometry but different statistics, the algorithm does not succeed in learning the underlying distribution to the same extent (Fig.~5B).

Our results demonstrate the capacity of a neural network to learn, at least partially, the structural relationship of ridges and valleys in a crease pattern of crumpled sheets. The next step is to understand the network's decision process, with the aim of uncovering the physical principles responsible for the observed structure. However, while interpretation of trained weights is currently a heavily researched topic, see~\cite[among many others]{Smilkov2016,Frosst2017,Sundararajan2017}, there is not yet a standard method to do so. Our ongoing work seeks to probe the network's inner workings by perturbing the input data. For example, we can individually alter input pixels and quantify the effect of perturbation on the prediction relative to the original target. Alternatively, we can examine the effect of adding or removing creases, or test the prediction on inputs that do not occur naturally in crumpled sheets. Some preliminary results are discussed in Sec.~IV of the SI.

Improving the experimental dataset by performing dedicated experiments, or replacing the simulated flat-folding with simulated crumpling data are also promising future directions. While we have only demonstrated the advantages of data augmentation for one problem, it is tempting to imagine how it may apply to other systems in experimental physics. In addition to providing insights into the structure of crease patterns, a quantitative predictive model (i.e.~an oracle) could serve as an important experimental tool that allows for targeted experiments, especially when experiments are costly or difficult. As shown above, a trained neural network is able to shed light on where order exists, even if the source of the order is not apparent. 

Replacing the scientific discovery process with an automated procedure is risky. Frequently hypotheses which were initially proposed are not the focal points of the final works they germinated, as observations and insights along the way sculpt the research towards its final state. This serendipitous aspect of discovery has been of immense importance to the sciences and is difficult to include in automated data exploration methods, which is an area of ongoing research~\cite{Raccuglia2016,GooglePlasma,Ren2018}. By showing that data-driven techniques are able to make non-trivial predictions on complicated systems, even in a severely data-limited regime, we hope to demonstrate that these tools should become a valuable tool for experimentalists in many different fields.

\section*{Materials and Methods}
\paragraph*{Experiments} Experimental flat-folding and crumpling data were performed on $10 \mathrm{~cm} \times 10 \mathrm{~cm}$ sheets of 0.05~mm thick Mylar. Flat folds were performed successively at random, without allowing the paper to unfold between successive creases. Crumpled sheets were obtained by first rolling the sheet into a 3~cm diameter cylinder and then applying axial compression to a specified depth between 7.5~mm and 55~mm. Sheets were successively crumpled between 4 and 24 times.

To image the experimental crease network, crumpled/flat-folded sheets were opened up and their height profile was scanned using a home-built laser profilometer. The mean curvature map was calculated by differentiating the height profile, and then denoised using a custom Radon-based denoiser, the implementation details of which are given in Sec.~I of the SI. A total of 506 scans were collected from 31 different experiments.

\paragraph*{Network architecture and training}
Data was fed into a fully convolutional network, based on the SegNet architecture~\cite{segnet2015} with the final soft-max layer removed, as we did not perform a classification problem. The depth of the network allows for long-range interactions to be incorporated without fully connected layers. The network was implemented in Mathematica and optimization was performed using the ADAM optimizer~\cite{Kingma2015} on a Tesla 40c GPU with 256~GB of RAM and a computer with a Titan V GPU and 128 GB of RAM. Code is freely available.

For training, the \emph{in silico} generated input data was  augmented with standard data-augmentation methods: symmetric copies of each original were generated by reflection and rotation. All images were down-sampled to have dimensions of $224 \times 224$ pixels. For crumpling data, creases were also linearized to look more similar to the experimental input. An example of the effect of linearizing is shown in Fig.~S2 of the SI.

\paragraph*{Noise}
Noise was added to the input in a few different ways, presented In Fig.~3B. The noise of each panel was generated as follows:
\begin{enumerate}[A.]
\item No noise.
\item ``White'' noise: Each pixel was randomly toggled with 5\% probability.
\item Random Blur: Input was convolved with a Gaussian with a width drawn uniformly between 0 and 3. The array was then thresholded at 0.1. Here and below ``thresholded at $z$'' means a pointwise threshold was imposed on the array, such that values smaller than $z$ were set to 0 and otherwise set to 1.
\item Each pixel was randomly toggled with 1\% probability, then passed through Random Blur (C).
\item Input was Random Blurred (as (C)) but thresholded at 0.55.  We denote the blurred-and-thresholded input as $\tilde X$. Then, $\tilde X$ was noised using both additive and multiplicative noise, as follows: $Y$ and $Z$ are two random fields drawn from a pointwise uniform distribution between 0 and 1 and convolved with a Gaussian of width seven (pixels) and thresholded at 0.55. Finally, the ``noised'' input is
$$
\min\big(\tilde X + (1-Y),\,1\big) ( 1 - Z)\ .
$$
\end{enumerate}

\section*{Acknowledgments}
We thank an anonymous referee for this suggestion to perturb the \emph{in-silico} input data. This work was supported by the National Science Foundation through the Harvard Materials Research Science and Engineering Center (DMR-1420570). SMR acknowledges support from the Alfred P. Sloan research foundation. The GPU computing unit was obtained through the NVIDIA GPU grant program. JH was supported by a Computational Science Graduate Fellowship (DOE CSGF). YBS was supported by the JSMF post-doctoral fellowship for the study of complex systems. CHR was partially supported by the Director, Office of Science, Computational and Technology Research, U.~S.\ Department of Energy under Contract No.\ DE-AC02-05CH11231.

\paragraph*{Data availability} Source code is available at \url{https://github.com/hoffmannjordan/Crumpling}. All data needed to evaluate the conclusions in the paper are present in the paper and/or the Supplementary Materials. Additional data available from authors upon request.


\onecolumngrid
\vspace{4cm}
\begin{center}
	\textbf{\large Supplemental Materials}
	\vspace{1cm}
\end{center}
\twocolumngrid

\setcounter{equation}{0}
\setcounter{figure}{0}
\setcounter{section}{0}
\setcounter{table}{0}
\setcounter{page}{1}
\makeatletter
\renewcommand{\theequation}{S\arabic{equation}}
\renewcommand{\thefigure}{S\arabic{figure}}
\renewcommand{\thesection}{S-\Roman{section}}
\renewcommand*{\thepage}{S\arabic{page}}

\section{Radon transform based detection method}
Here we detail the detection method used to identify crease networks from maps of mean curvature prior to machine learning. We refer to our technique as a Radon-based detection method, as it repurposes the key principle behind a Radon transform---recovering a signal through integration along directed paths---for crease detection. By integrating a quantity of interest, in our case the mean curvature, along paths of regularly spaced orientations within local regions of the curvature map, we construct a signal array that enhances the signature of creases and reduces noise. A strong signal is recovered if an integration path coincides with the direction of an extended structure such as a crease; a weak signal is produced by features that are point-like or isotropic, representative of noise and fine texture in the data. The raw curvature maps of each $10\textrm{~cm}\times10\textrm{~cm}$ sheet are $3000\times3000$ pixels. Prior to processing, curvature maps are downsampled for computational efficiency. A downsampling factor of 4 was found to preserve the integrity of the crease pattern while providing a useful speedup in computation for a final resolution of $75$ pixels per cm. Next, a linear integration path is centered about a given pixel of the curvature map, traversing the diameter of a fixed circular local window. The average curvature along a particular direction is computed by exact numerical integration of the bicubic interpolant on the grid defined by pixel centers. The integration direction is systematically rotated about the central pixel, and the maximum average curvature over all path orientations is selected as the signal. This process is repeated for all pixels in the curvature map, resulting in a signal array of only the average curvatures that are a maximum along local, linear paths. Integrals along 24 equally spaced path orientations on the interval of 0 to 180 degrees were considered at each pixel and the maximum selected as the signal. We examined a range of integration path lengths up to 8~mm, as the integration window defines a length scale that must accommodate features of varied sizes. While smaller integration paths can detect finer details particularly at low crease densities, they sacrifice some of the advantage afforded by longer paths in accruing a strong signal that is well separated from noise. An integration path length of 3.2~mm suitably mediated such effects and provided a clear crease network. Finally, global and local thresholds are applied to the signal array to separate the real creases from the background noise. A combination of the two was observed to work well in retaining the desired crease network: The global threshold is more permissive of noise but acts uniformly across the signal array, while the local threshold accommodates variations in signal intensity, and thus provides sensitivity to softer (less sharp) creases. We use a global threshold of $0.12$ as the minimum signal intensity retained as a crease (0.12 is approximately 10\% the magnitude of the largest creases), and set the local threshold to label as noise any pixel whose intensity falls below $20\%$ of the maximum signal in a $3.2\textrm{~mm}\times3.2\textrm{~mm}$ neighborhood centered about the pixel. In training with crumpled sheets, the crease networks were also linearized as shown in Fig.~\ref{fig:graph}. This was done with a custom script that skeletonized the input and used the \emph{Mathematica} function \texttt{MorphologicalGraph}.

\begin{figure}
	\centering
	\includegraphics[width=0.98\columnwidth]{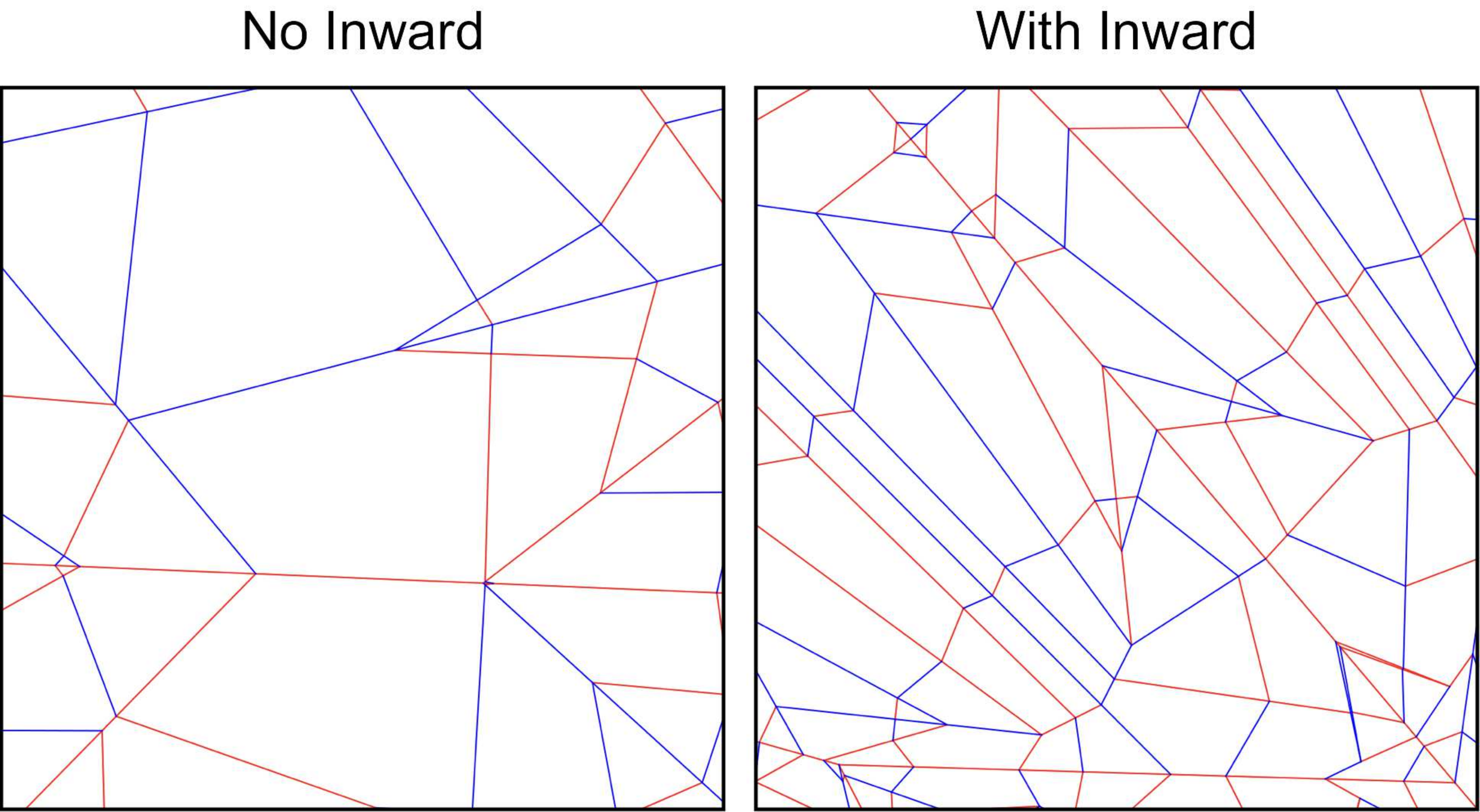}
	\caption{\label{fig:examples} \textbf{In-silico generated flat-folded crease networks}. Two random flat-folding patterns, with (left) and without (right) inward folding. Ridge folds are colored red and valley folds are colored in blue.}
\end{figure}

\section{\emph{In silico} generation of flat-folding data}
A custom code was written in C++ to simulate flat folding. The code makes use of the \textsc{Voro++} software library~[30], which provides routines for fast manipulation of polygons. To begin, the sheet is represented as a single square. To simulate a simple flat fold on a given chord, the square is cut into two polygons, and one polygon is reflected about the chord. Subsequent flat folds are simulated by taking the collection of polygons representing the folded sheet, cutting them by a given chord, and reflecting those on one side about the chord. Throughout the process, each polygon keeps track of an affine transformation from its current configuration back to its position in the original square sheet. By transforming all polygons back to the original sheet, the flat folding map of valleys and ridges can be constructed. The code can also simulate inward folds where a ray is selected and the sheet is pinched in along this ray. For computational efficiency, the code computes a bounding circle during the folding process, whereby the collection of polygons representing the folded sheet is wholly within the circle.

While folding along a given chord is strictly well-defined, there is no natural way to draw a random chord from a distribution (e.g.~Bertrand's paradox in probability theory) and a choice must be made regarding the way a chord is drawn. Our choice is the following: A fold is determined by a straight line in $\mathbb{R}^2$ and therefore can be parameterized by its angle and offset. At each iteration the angle is drawn uniformly in the range $[0,2\pi)$ radians and the offset uniformly over the bounding circle. If the chosen fold line does not actually create a fold (because the line misses all polygons) then a new angle and displacement are chosen and so forth. For inward folds, we first choose a point uniformly inside the bounding circle, then determine if the point is inside any polygon; if not, we keep choosing new points until we find one that is. We then choose a random orientation for the ray from this point and two random angles $\alpha$ and $\beta$ uniformly from $(0,\pi)$ for the first two folded segments that are counter-clockwise from the ray, after which the remaining two angles at the point are given by $\gamma=\pi-\alpha$ and $\delta=\pi-\beta$.

\begin{figure}[t]
	\centering
	\includegraphics[width=\columnwidth]{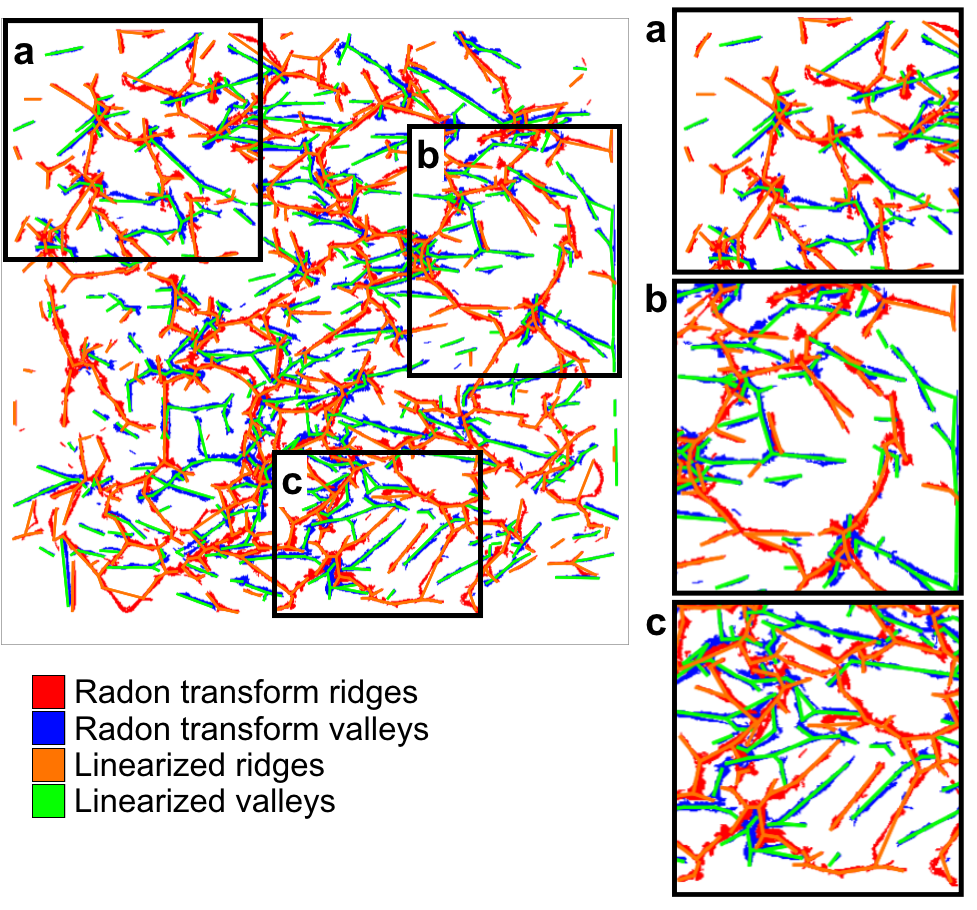}
	\caption{\label{fig:graph} \textbf{Comparison between the preprocessed curvature map and the linearized version.} The denoised curvature map of an entire crumpled sheet with three enlarged insets (a-c) for better visibility. Red and blue are creases retained after the Radon-based denoising,  green and orange are the linearized representation.}
\end{figure}

Our data set was generated by folding the sheet $n$ times, where $n$ is chosen uniformly in the range from 7 to 15. Each fold has a random sign (ridge or valley) with equal probability. For each sheet, the probability of inward folds was chosen uniformly over the range $[0\%,50\%]$. Figure~\ref{fig:examples} shows a selection of generated crease patterns.

\begin{figure*}[hbt]
\centering
\includegraphics[width=\textwidth]{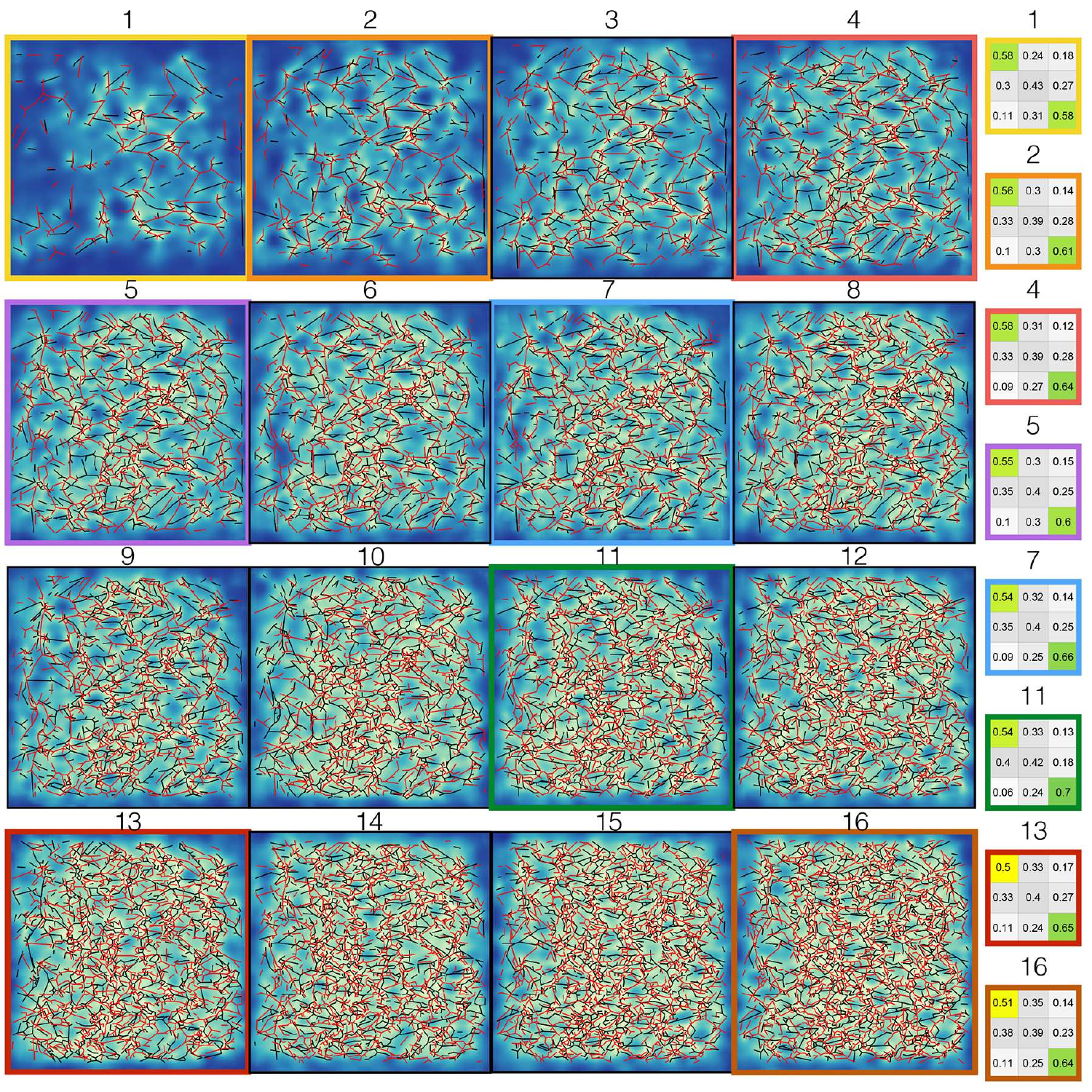}
\caption{\label{fig:all} \textbf{Prediction on a sheet that was crumpled 16 times}. The prediction is shown in blue for a given set of valleys (black). The true creases are overlaid in red .Confusion matrices for 8 of the 16 matrices are shown in the right. The color corresponds to the outline of the matrix.}
\end{figure*}

\begin{figure*}[hbt]
\centering
\includegraphics[width=\textwidth]{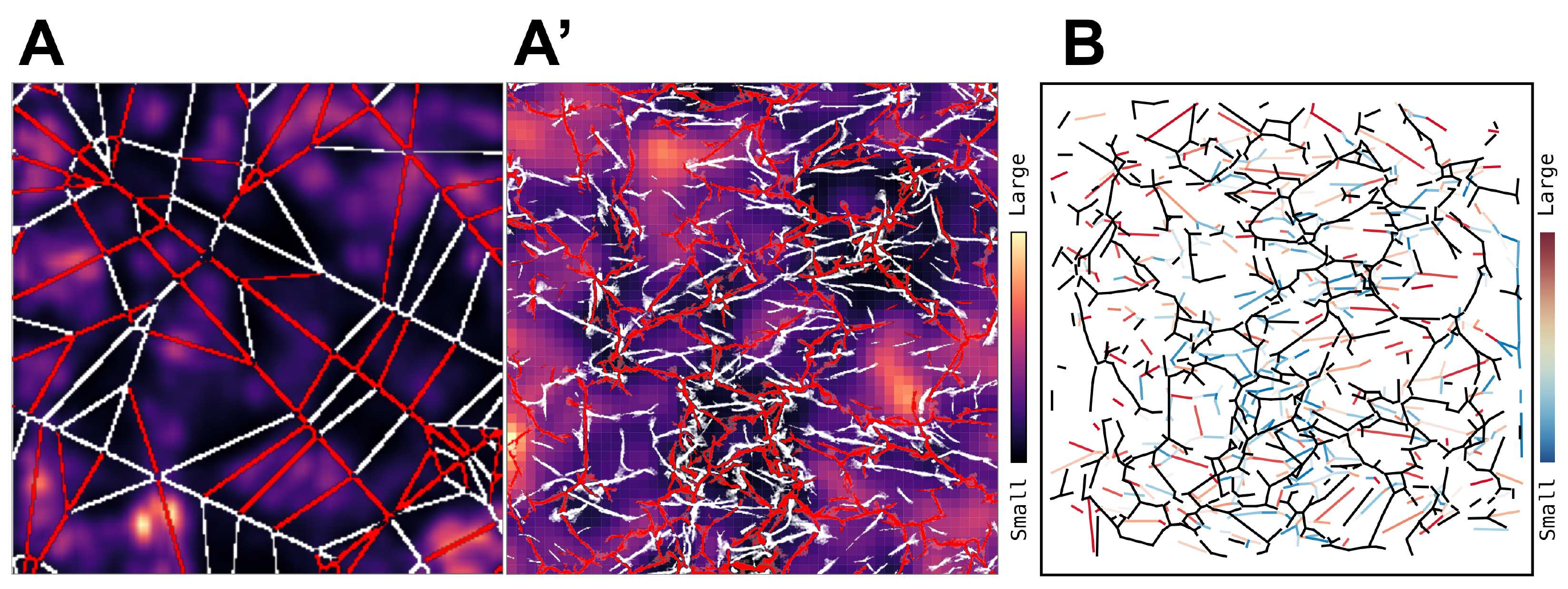}
\caption{\label{fig:perturbations} \textbf{Additional test results.} \textbf {A} The result of approximate differentiation (see text in Sect.~\ref{sec:probing}) on flat-fold (\textbf{A}) and crumpling (\textbf{A'}) inputs. Unfortunately, experimentally testing these results or correlating them with other physical quantities proved difficult. \textbf{B} Creases colored by the magnitude of the change caused by their removal (Eq.~\ref{eq:change_magnitude}). Cooler colors correspond to weaker change and warmer colors to stronger change. While some trends are clearly discernible (e.g.~there is a strong correlation between the change magnitude and the crease length), we are still trying to interpret these results in terms of the underlying physics.}
\end{figure*}

\section{Prediction on 16 sheets}
The validation set (an experiment held out from training) consists of 17 successive crumples of the same sheet of paper. In Fig.~\ref{fig:all}, we show the prediction on the first sixteen of these sheets. For each prediction of an entire sheet, the image was computed in overlapping patches of size $224 \times 224$. Each pixel was considered to be the average value based on a sequence of predictions. Preliminary work was done on automatically detecting regions that were the best and the worst predicted. This along with aspects discussed below, are the topic of ongoing work.

\begin{figure}[h]
	\centering
	\includegraphics[width=\columnwidth]{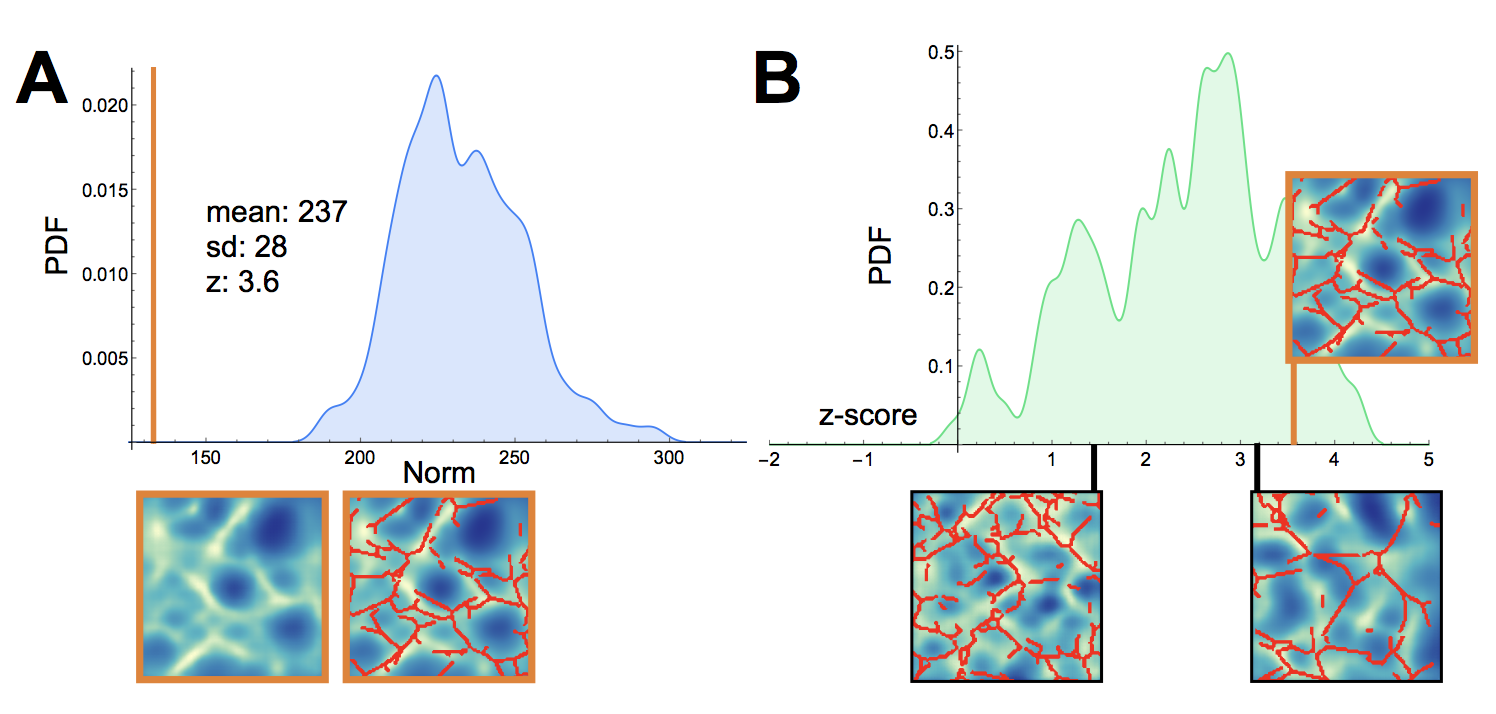}
	\caption{\textbf{Prediction Accuracy.}  \textbf{A} The loss (orange line) of a given reconstruction (bottom) compared to the of losses distribution from all other patches from similarly crumpled sheets. From this data we calculate the $z$-score of this patch to be 3.6. \textbf{B} Repeating this procedure for all patches, we calculate the distribution of $z$-scores, giving an average $z$-score of nearly 3. Three representative patches are shown at their $z$-score location.}
	\label{fig:error} 
\end{figure}

\section{Probing the network: Ongoing work} \label{sec:probing}
In their paper, Lehman \emph{et al.}\ discuss some computational oddities in the field of computational evolution~[39]. They present a series of important ideas through short tales where the computer produced unexpected behavior that, when understood, were a key step in learning how to successfully use the computational tools. We think examples similar to these are important to share as the use of machine learning in the experimental sciences is still in its infancy. In this spirit, we discuss some of our attempts to tie the predictions of the network back to the underlying physics of crumpling.

A potential pitfall of using neural networks is that they will provide an output for any input, no matter how absurd either the input or output is. No warning appears. This is powerful, but requires caution, as neural networks allow for predictions on inputs that are physically impossible to create. Thus, one should take all the following probing attempts with a grain of salt.

It is tempting to ``differentiate'' the input signal to see if perturbations at any particular location cause large changes in the neural network's prediction. In Fig.~\ref{fig:perturbations} A and A' we do this, perturbing the input (empty space and white lines) by making each pixel slightly more crease-like if it is not a crease or less crease-like if it is a crease. The background color shown is the magnitude of the change relative to the original prediction. Our hope was that this map may correlate with some known aspect of the physics. However, we do not think that this is the case. We tried aligning sequential images and estimating whether new creases tend to form with higher probability in regions that correlate with this sensitivity map---we do not find this to be the case. We are currently exploring more sophisticated ways of differentiating the trained network.

Similarly, we can ask questions such as: What would happen if we translate a particular crease 5~mm to the left? What if we artificially remove parts of folds in flat-folding? What if we remove entire creases? In Fig.~\ref{fig:perturbations}B, we present the results of removing entire creases form a crumpled sheet. The ridges are colored by their total effect on the prediction, that is, if the original input is $X$ and the perturbed input is $\tilde X$ (after removing a crease), we define the total magnitude of the change as
\begin{equation}
\sum_i\Big(\mathcal{N}(X)_i-\mathcal{N}(\tilde X)_i\Big)^2\ ,
\label{eq:change_magnitude}
\end{equation}
where $i$ runs over all pixels. The hope is that these unphysical perturbations to the input can provide insight into the working of the network. However, as stated above, interpreting them should be done with care, since in these cases the input to the neural net might be too dissimilar to anything in the training data, making predictions less reliable.

\section{Another approach to error quantification}
It is common to benchmark Machine Learning prediction accuracies with respect to a suitably-defined random guess. For example, in the \texttt{MNIST} digit recognition task, making random choices will achieve a 10\% accuracy, because there are only ten classes to choose from. In our case, however, there exists no generative model for crease networks, so there is no random guess that we can compare the output of our network to. As a surrogate, we can draw a random crease network from our data. That is, we compare our predictions on a given patch to many patches from other, similar experiments. This is presented in Fig.~\ref{fig:error}: For a given patch, we compute the loss of our prediction (Eq.~1 in the main text) compared to the true value. In Fig.~\ref{fig:error}A we compare this loss with the distribution of losses obtained by comparing other patches to the true value. Examining hundreds of different predictions, in Fig.~\ref{fig:error}B, we find that our predictions have an average $z$-score of nearly 3. The $z$-score for a patch is defined as $z=(\mu - L)/\sigma$ where $L$ is the loss for this patch and $\mu$, $\sigma$ are, respectively, the mean and standard deviation of all losses calculated from other patches on the same true value.
We find that the prediction returned by the net is substantially better than patches taken from other experiments.

Additionally, we can compute the Pearson correlation between the distance transform of the input and target, as well as between our prediction and the target. In the following table we show, for four representative crumple iterations, the Pearson correlation between the target distance map and either the distance map of the input or the network prediction.\\[3mm]
\begin{tabularx}{\columnwidth}{p{2cm}Xp{25mm}}
  Iteration &  Input distance map & Prediction \\
  \hline
 1 & 0.44 & 0.68 \\
 3 & 0.35 & 0.66 \\
 6 & 0.31 & 0.54 \\
 11 & 0.52 & 0.74 \\[3mm]\end{tabularx}
It is seen that our prediction is significantly better than simply returning the distance transform of the input.

\section{Perturbing the \emph{in silico} data}
\label{sec:perturb}

\begin{figure}[b]
	\centering
	\includegraphics[width=\columnwidth]{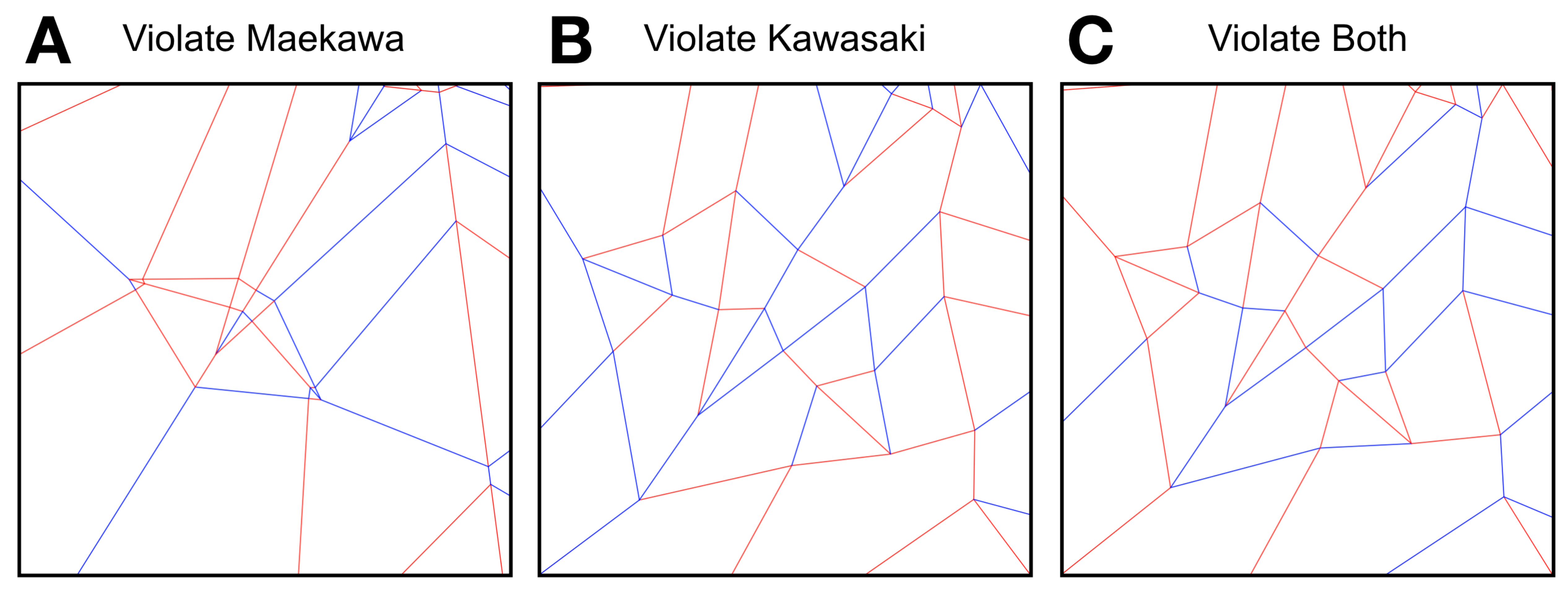}
	\caption{\textbf{Examples of perturbed in silico data.} \textbf {A-C} One realization of each perturbed in silico data set, corresponding to the perturbations described in Sec.~\ref{sec:perturb}. Code to generate all types of perturbation available online.}
	\label{fig:violate}
\end{figure}

As discussed in the main text, we assessed the sensitivity of the prediction accuracy to perturbing the \emph{in silico} data. In Fig.~\ref{fig:violate} we present examples of perturbed crease networks (panels A-D) and the resulting validation loss as a function of the number of times the sheet was crumpled. It is seen that all perturbations lead to inferior predictions. 

Perturbations were performed in the following manner:
\begin{enumerate}
	\item Maekawa's theorem was violated by taking flat-folding networks and randomly reassigning curvatures (ridge/valley) to each crease. On average, Maekawa's rule is violated in 50\% of the vertices.
	\item Kawasaki's theorem was violated by perturbing the position of the vertices, while keeping the topology of the network fixed and ensuring that creases do not cross each other. This results in alternate angles that no longer sum to $\pi$. The average absolute deviation of  from $\pi$ is 0.4, amounting to $\sim13\%$ change. The code is available on GitHub. 
	\item Finally, both rules were violated by combining both procedures 1-2.
\end{enumerate}

\end{document}